# Emergence of species in evolutionary ''simulated annealing''.

Muyoung Heo, Louis Kang and Eugene I. Shakhnovich

Department of Chemistry and Chemical Biology, Harvard University,
12 Oxford St, Cambridge, MA
E-mail: eugene@belok.harvard.edu

**Abstract**

Which factors govern the evolution of mutation rates and emergence of species? Here, we address this question using a first principles model of life where population dynamics of asexual organisms is coupled to molecular properties and interactions of proteins encoded in their genomes. Simulating evolution of populations, we found that fitness increases in punctuated steps via epistatic events, leading to formation of stable and functionally interacting proteins. At low mutation rates, species – populations of organisms with identical genotypes - form, while at higher mutation rates, species are lost through delocalization in sequence space without an apparent loss of fitness. However, when mutation rate was a selectable trait, the population initially maintained high mutation rate until a high fitness level is reached, after which organisms with low mutation rates are gradually selected, with the population eventually reaching mutation rates comparable to those of modern DNA-based organisms. These results provide microscopic insights into the dynamic fitness landscape of asexual populations of unicellular organisms.

\body

The concept of species is central to Biology (*1*). In the pre-genome era species were defined through combination of observable phenotypic traits, most importantly the ability to interbreed – making it challenging to define species for population of asexual organisms. With the advance of genome sequencing the concept of species has undergone transformation to acquire a more detailed, microscopic meaning as a collection of organisms possessing (almost) identical genomes (though a precise definition of a specie,, is still a matter of a lively debate (*2*)). Beginning from Darwin's work, the origin of species has been at the center of studies in evolutionary biology. The existence of species is often postulated in theories of evolution through a ''single fitness peak'' assumption (*3*). However, it may be at variance with the observation that proteins and nucleic acids are mutationally robust (*4-6*). An approach which presents a distinct departure from the single fitness peak assumption has been proposed recently (*7*). There, the authors found a universal upper limit on the mutation rate, about six missense mutations per essential coding part of the genome per generation, beyond which essential proteins in an organism lose their stability and the population goes extinct (*7*). Several RNA viruses indeed have mutation rates close to the speed limit while even simplest DNA organisms, such as bacteria, have mutation rates typically 100-1000 fold below it (*8, 9*). Apparently mutation rate is a selectable trait, and there is an evolutionary pressure to bring it down way below the ''speed limit'', resulting in the genomic homogeneity of populations. However, the source of selective pressure to decrease mutation rates is not clear: while it is believed that low mutation rates may be advantageous against detrimental mutations (*10*), many bacteria can increase their mutation rates 10-100 fold without noticeable loss of fitness (*11-13*).

These fundamental issues should be addressed within an approach where population genetics and protein biophysics are realistically coupled in a microscopic model of organisms which does not assume a'priori a single fitness peak or any other form of single species domination.

Recently, we developed such a computational model and applied it to the study of early evolution (*14*). In this model, cells have genomes that encode simple coarse-grained model proteins. The proteins are modeled as 27-amino acid residue lattice polymers whose thermodynamic properties, including stability in their native state, can be exactly derived from their sequences (*15*). Despite successes of this model in describing possible early events in evolution where protein folds had been discovered, its focus on protein stability as a single genotypic trait may limit the ability of the model to reproduce biological events.

Here we study an ab initio microscopic evolutionary model which considers protein function and its effect on fitness. Specifically we consider simple organisms whose genomes carry a fixed number – 3 – of genes, and the protein products of these genes interact so that fitness of a cell depends not only on protein stability but also on interactions between proteins. In particular, we assume a simplest protein-protein interaction (PPI) network in our simulated cells where the product of gene 1 functions in a monomeric state, while products of gene 2 and gene 3 function as a tightly bound dimer (see Fig.1). This requirement is expressed via a genotype-phenotype relationship which

derives the fitness – growth rate – of model organisms from molecular properties of their proteins:

$$b = b_0 \frac{P_{nat}^1 P_{nat}^2 P_{nat}^3 F_1 F_{23} P_{int}^{23}}{1 + \alpha (C_1 + C_2 + C_3 - C_0)^2} \quad (1)$$

where $b_0$ is a base growth rate, $C_1, C_2, C_3$ are total production levels for proteins 1, 2, 3 respectively, $C_0$ is an optimal production level for proteins in a cell, and $\alpha$ is a control coefficient which sets the range of allowed deviations from optimal production levels. $P_{nat}^i$ is stability (the Boltzmann probability to be in the native state, see Supplementary text) for the protein product of gene $i$, $F_1$ is concentration of free protein 1, $F_{23}$ is concentration of complex between protein 2 and protein 3 determined using the Law of Mass Action (LMA) (see Supplementary text). $P_{int}^{23}$ is the Boltzmann probability for proteins 2 and 3 to be in a unique functional conformation out of 144 possible mutual orientations of the encounter complex between them (see Supplementary text). The biological meaning of the genotype-phenotype relationship given by Eq.(1) is simple: the numerator states that birth rate is proportional to the concentration of *functional* proteins while the denominator states that unlimited growth of protein production would hamper cell growth due to depletion of resources. Proteins can interact with each other in cellular cytoplasm and form both functional and non-functional pair-wise complexes (see Fig.1).

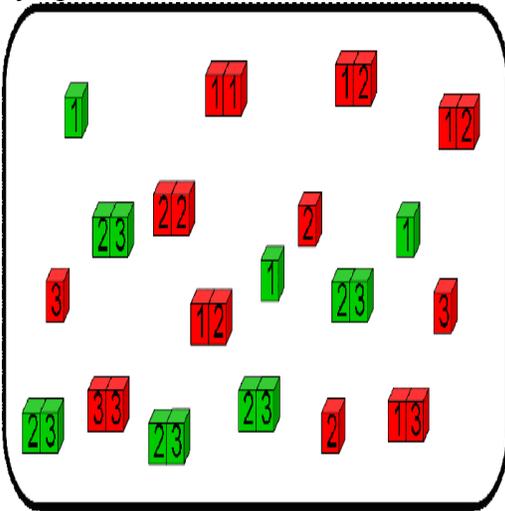

**Figure 1**.
*A schematic diagram of the model. A model organism is depicted to have 3 types of genes, which are expressed into multiple copies of model proteins. Proteins can stay as monomers or form dimer conformations, whose concentrations are determined by interaction energies among proteins and Law of Mass Action equations. Green and red cubes respectively represent favorable and unfavorable states of interaction.*

We simulated evolution of populations of model organisms for various mutation rates in the range 0.0001-0.05 mutations per gene per replication (see Supplementary Text). In order to determine which results depend on protein function (PPI in our model) we carried out, in parallel, control simulations where the fitness – birth rate – depended only on the product of stabilities of the three proteins in Eq.(1) but not on their interactions (see Supplementary text).

Fig.2 shows the population dynamics for typical evolution runs of the model for low mutation rate m=0.001 (A), for high mutation rate m=0.05 (B) and control simulations where PPI are excluded (C). Evolution proceeds in punctuated steps to discover viable organisms with dramatically higher growth rates (middle panels). In contrast, in the control simulation (Fig.2C) the birth rate increases insignificantly. Importantly, organisms evolving at higher mutation rates achieve higher fitness than those evolving at lower mutation rates. While final fitness varies significantly between evolution runs, distributions of fitness over evolution runs are dramatically biased toward higher fitness values for higher mutation rates. Next, we address the emergence of species. We define a ''specie'' as a group of organisms that possess identical protein sequences. The dominant specie in the population is the one that has the most organisms. Middle panels of Figure 2 show how the fraction of the dominant specie in the population changes with time. Dramatic difference between low mutation rate and high mutation rate cases is apparent: while species do exist at low mutation rate – the dominant specie constitutes more than 80% of the population – they quickly disappear at higher mutation rate (Fig.2B). Also we note that the ''oscillations'' in the dominant species fraction at low mutation rate, Fig.2A, correspond to the emergence of new species due to fixation of new mutations in the population (see below). Apparently these new mutations are beneficial as their fixation occurs concurrently with punctuated increases in fitness. One can argue that such dramatically different behavior with respect to the formation of species may be due to the trivial fact that at low mutation rate, genomes do not acquire sufficient number of mutations to diverge from initial sequences. To this end, we compare the result of the full model at low mutation rate – $m = 0.001$ mutations per gene per replication – with the control model where fitness depends on stability of proteins but not on PPI. The control simulations are run at conditions where effective mutation rate, $m \cdot b$, is exactly the same as it is for the low $m$ full model simulation of Fig.2A. We can see that in the control simulation no species are formed – suggesting that complex (functional and non-functional) interactions between proteins in this model are responsible for the formation of species.

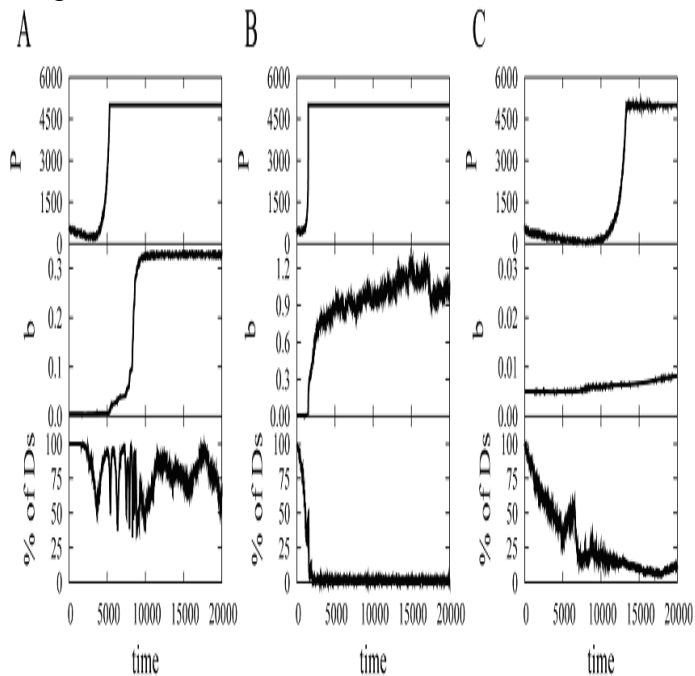

**Figure 2**.
*Population dynamics of the model I. Panels in each figure show population (P), birth rate (b) and fraction of dominant species (% of Ds) of organisms. (A) At low mutation rate $m=0.001$ (mutations per gene per replication), the emergence of a beneficial strain restores the fraction of dominant species and sequence entropies (Figure 3). (B) At high mutation rate $m=0.05$, the fraction of dominant species saturates around 0.012 because new beneficial strains of the model organism emerge before the former ones get fixed – a clonal interference phenomenon. (C) The results of control simulations with the same effective mutation rate $m \cdot b=0.000327$ as (A) are shown, demonstrating that the formation of species does not emerge.*

In order to better understand how new species are formed in evolution, we separately consider mutations in all three proteins. Fig.3 shows the evolutionary time dependence of sequence entropy S(p) for each of the three proteins. This quantity is obtained from alignment of all sequences in the population for each of the proteins (See Methods for definition). Low S indicates that all proteins of a given locus in the population have very similar sequences while high S suggests substantial sequence heterogeneity in the population. Fig.3 shows that in most cases beneficial mutations are acquired as a result of epistatic events when mutations are fixed ''synchronously'' in several proteins.

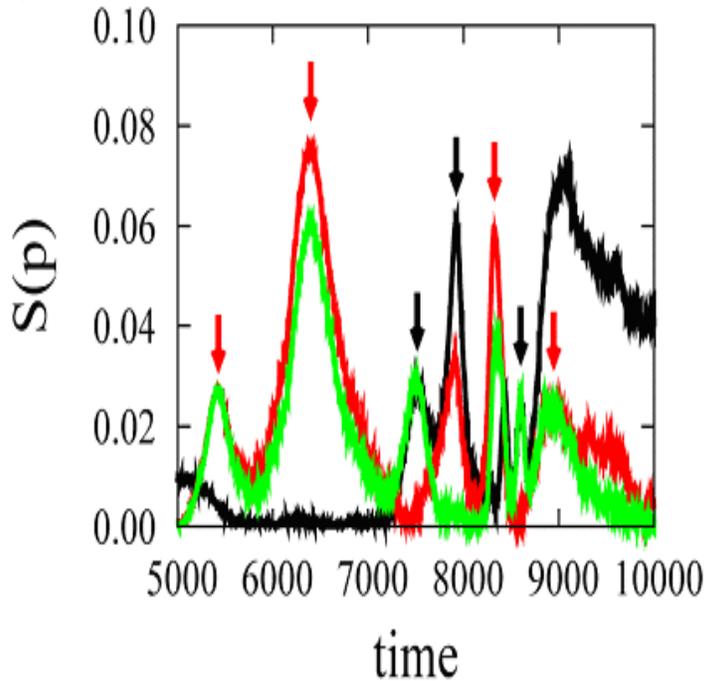

**Figure 3**.
*Sequence entropies of proteins at m=0.001. Sequence entropies (S(p)) of protein p=1 (black), 2 (red) and 3 (green) are calculated from the sequence alignment of each protein of all organisms in the system. Epistatic events between gene 2 and 3 (marked by red arrows) and those involving gene 1 (marked by black arrows) are observed.*

A deeper insight into microscopic mechanisms of evolution in this model can be gained from analyzing molecular properties of evolved proteins (Fig.4). Evolution of a tightly bound functional complex between proteins 2 and 3 proceeds in punctuated steps at which specific mutations on surfaces of both interacting proteins get fixed both at higher and lower mutation rates. Another functionally important property which definitely evolves in discrete steps is the fraction of monomeric protein 1. These mutations decrease surface hydrophobicity of this protein, making it less prone to non-functional interactions with itself and other proteins.

The bottom panel of Fig.4 shows evolution of normalized pairwise sequence identities for all three proteins. At lower mutation rate, sequences of functionally interacting proteins 2 and 3 diverge slower than those of functionally ''unrelated'' proteins 1, 2 and 1, 3. This is consistent with the proposal that interacting proteins evolve more slowly (*16*). At higher mutation rate, (Fig.4A) this effect disappears.

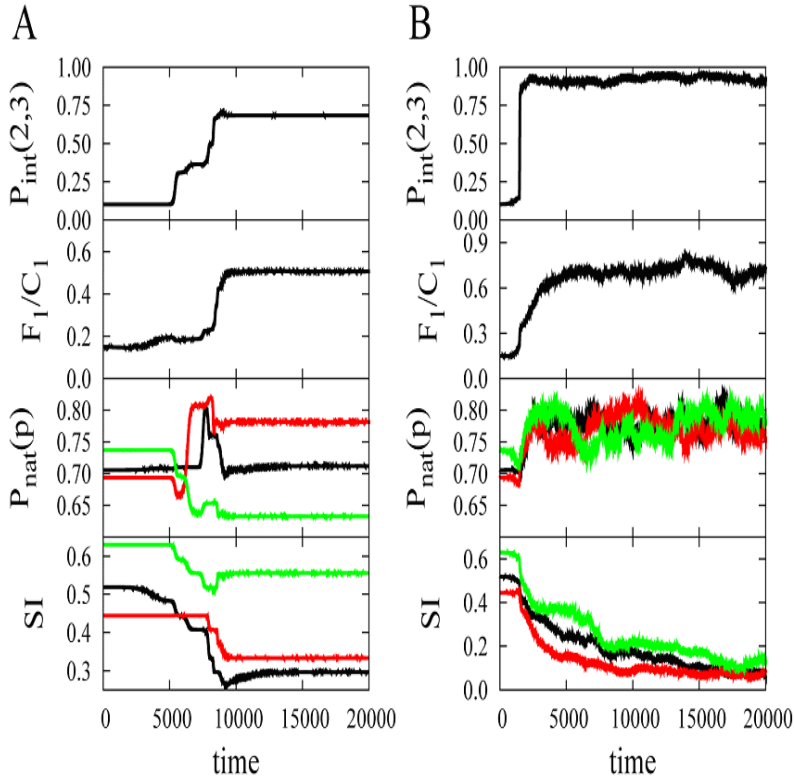

**Figure 4**. *Population dynamics of the model II. The microscopic variables of interaction probability between protein 2 and 3 ($P_{int}(2,3)$), the fraction of monomer concentration of protein 1 ($F_1/C_1$), the stabilities ($P_{nat}(p)$) of protein 1 (black), 2 (red) and 3 (green) and sequence identities (SI) between protein pairs 1-2 (black), 1-3 (red) and 2-3 (green) are shown from top to bottom in these panels. (A) At low mutation rate m=0.001, The birth rate (Figure 2A) increases through discrete steps, and each step corresponds to the increase of $P_{int}(2,3)$, $F_1/C_1$, or $P_{nat}(p)$. (B) At high mutation rate m=0.05, the birth rate (Figure 2B) grows continuously after t=1500, and no stepwise increase exists in the $P_{int}(2,3)$, $F_1/C_1$, and $P_{nat}(p)$ plots – again due to the effect of clonal interference.*

We performed evolutionary simulations in a broad range of mutation rates and the results are summarized in Fig.5. It is quite clear from Fig.5 that organisms evolve to higher fitness at higher mutation rates and the probability for the population to survive rapidly approaches 1 as m increases. Fitness of the organisms, on average, increases at higher mutation rate (see Fig.5A). Fig.5B shows that species are lost as mutation rate increases – at higher mutation rate, populations represent collections of organisms with vastly divergent genomes, yet higher mutation rates allow them to achieve, on average, higher fitness.

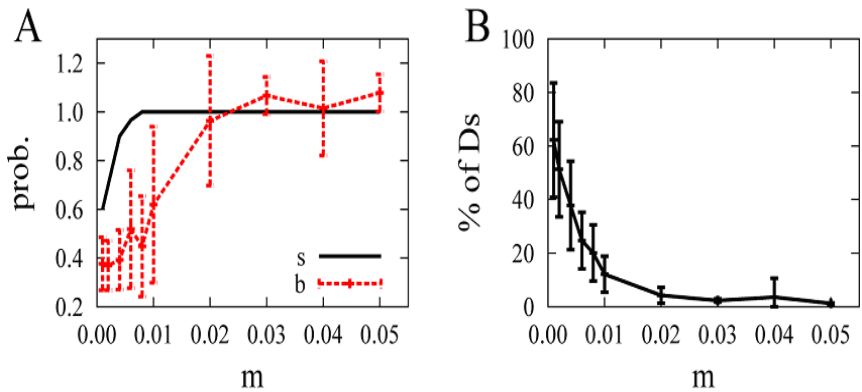

**Figure 5**. *The dependence of population dynamics on mutation rate. (A) Black solid and red dashed lines show the survival probability of the population and*

*average birth rate of the model organisms in the population at t=20000 in 30 runs as a function of mutation rate. Rapid change of the birth rate occurs between m=0.01 and 0.02. (B) The fraction of dominant species drops below 10% after the transition. The error bars show variation between runs.*

This finding presents an interesting dilemma. Indeed, at mutation rates which are comparable to those in modern bacteria, stable species do exist in our model. On the other hand, at much (100-500 fold) higher mutation rates species disappear yet fitness is not lost – rather it appears that higher mutation rates are beneficial. In order to address this issue, we changed the model slightly, now allowing mutation rates to fluctuate (see Supplementary Text). In this modification of the model, the mutation rate itself is a selectable trait and the question that we ask is whether particular mutation rate(s) will be selected in evolution.

The results of simulations are shown in Fig.6, which presents a very long (400,000 time steps) run. It appears that mutation rate is indeed a highly selectable trait: When simulations start from an initially high mutation rate (the case $m_{init} = 0.05$ per gene per replication is shown in Fig.6) it stays at this level for some time as system evolves to a high fitness level. However as fitness appears to plateau, the mutation rate starts to steadily decrease, dropping approximately 200 fold over the course of the simulation! The evolution of species is quite remarkable as well. At the initial stage while the population maintained the high initial mutation rate, species quickly disappeared concurrently with population and fitness growth – consistent with the picture for the constant mutation rate model. However, as sufficient fitness had been acquired and mutation rate started to slow down, the species structure of the population began to emerge. As seen in the two middle panels, both the fraction of dominant species in the population starts to grow and sequence entropy (for the whole genome, reflecting sequence diversity of the organisms in the population) starts to decrease, eventually reaching levels typical of the low mutation rate scenario shown in Fig.2A. However, in contrast to the constant low mutation rate case shown in Fig.2A, the variable mutation rate evolution resulted in species of much higher fitness with low mutation rate, a feat not achievable at constant low mutation rate. This behavior is observed in the vast majority of runs.

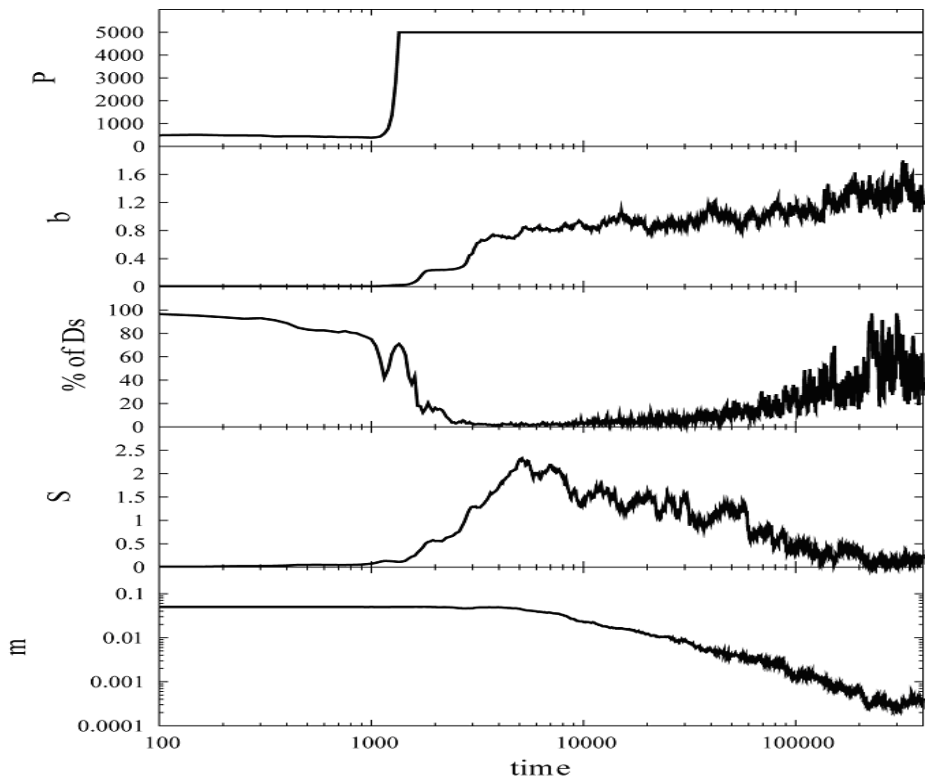

**Figure 6.** *Long-term population dynamics of the model with variable mutation*

*rates. The panels show population (P), birth rate (b), fraction of dominant species (% of Ds), combined sequence entropy of all three proteins (S), and mutation rate (m). The simulation starts at t=0 with mutation rate m=0.05. The system converges to low mutation rate by the end of the simulation at t=400000.*

Comparison of several independent evolutionary runs highlights important features of the fitness landscape. Starting from same initial sequences, evolutionary runs may arrive at different fitness levels (growth rate b). A more important question is whether similar long-time fitness values imply that evolution arrived at similar solutions, i.e. generated the same species. The analysis of the data gives a negative answer to this question – species that evolved in different evolutionary runs have very little genotypic similarity, representing independent evolutionary discoveries of high fitness populations of organisms.

## Discussion

Our model of evolution, despite its simplicity, represents, in many respects, a pretty dramatic departure from traditional phenomenological approaches. In these earlier models, a certain genotype-phenotype relationship – a ''fitness landscape'' such as a single fitness peak (*17*) or dominance of a single RNA structure (*18-20*) – is assumed a'priori and simulations and mathematical models are developed to explore consequences of these assumptions both on population and genomic levels. In contrast, here we do not a'priori assume any direct relationship between genome sequence and/or protein/RNA structure and fitness. Rather we posit that the growth rate of an organism is proportional to the concentration of functional proteins – in native monomeric form for the product of gene 1 and dimeric form between native products of gene 2 and 3. Further, we state that all proteins in a cell can interact, forming either functional or non-functional complexes. These straightforward physical assumptions directly relate an organism's fitness to physical-chemical properties of its proteins, such as their stability and propensity to form functional and non-functional complexes. The latter, in turn, are determined by protein sequences in a microscopic biophysically realistic model.

We observe very rich and biologically realistic behavior. We found that organisms evolve potent proteins which are both stable and capable of participating in functional PPI. In many cases, changes of fitness occur in a punctuated stepwise manner where a new beneficial mutation appears and then gets fixed in the population. This is reflected in the oscillation-like behavior in the fraction of dominant species seen in Fig.2 – for a certain time span both less fit and more fit (mutant) species coexist, after which an advantageous mutation gets fixed in the population. Another important observation is that fitness increases often occur through epistatic events when beneficial mutations get fixed simultaneously in several proteins. It is not very surprising that beneficial mutations get fixed simultaneously in interacting proteins 2 and 3. However the product of gene 1, which functions in the monomeric form, also participates in epistatic events. The reason for that observation is that our model takes into account all interactions between proteins – both functional and non-functional ones (*21*) – so participation of protein 1 in non-functional interactions affects its and its partners ability to stay in functional form. To this end, a beneficial mutation in protein 1 would minimize its non-functional interactions by making itself more soluble (data not shown).

Our simulations reveal a complex fitness landscape in the model. Indeed different runs produce species of similar fitness, yet their genome and proteome sequences show little or no homology. Furthermore, at low mutation rates different runs arrive at species of substantially different fitnesses, while at higher mutation rates fitness appears to converge to similar values in different runs. The evolution of fitness at higher mutation rates has a characteristic pattern of growth: initial rapid and huge (50-100 fold) increase in fitness is followed by its extended slower, more incremental growth. Similar behavior was found in experimental study of evolution of rapidly mutating RNA viruses (*22*). This is in contrast to the low mutation rate case where fitness increases slowly for first 1000 generations then plateaus after a few punctuated jumps and stays unchanged for the rest of the run, reaching an adaptive peak – similar to what was observed in E.coli evolution experiments (*23, 24*). Clearly these observations suggest that there are numerous multiple-maxima in the fitness landscape and the population ''freezes'' in the adaptive peak which is reached first. Subsequent increase of fitness becomes less likely as it requires overcoming fitness ''barriers'' in sequence space, probably via correlated mutations occurring in several proteins.

A key finding of our study is the emergence (and disappearance) of species – ensembles of organisms which carry identical genomes. Species are observed in our model only at low mutation rates (*25*) – as mutation rate gets higher, the organisms get delocalized in sequence space, and species disappear (see Fig.4). This is reminiscent of the ''error catastrophe'' phenomenon predicted by Eigen within a single fitness peak quasispecies approach (*17*). However, in stark contrast to Eigen's prediction, in our model higher mutation rates confer higher fitness, despite sequence space delocalization. It was argued that experimentally observed decay of viral titer at high mutation rates provides experimental support for the error catastrophe prediction (*26*). However, the decay of viral populations at elevated mutation rate is likely caused by a different phenomenon – mutational meltdown when viral proteins acquire mutations which make them unstable (*7*). The mutation rates studied here are much lower than the mutational meltdown threshold, which is approximately six mutations per genome per generation (*7*).

Higher fitness at elevated mutation rates is due to the fact that beneficial mutations are more readily available, or in other words, sequence space is more thoroughly searched at high mutation rate. Indeed many RNA viruses operate close to the ''mutational speed limit'' (*7-9, 27, 28*). However, DNA-based organisms have much lower mutation rates, and they form species. How can one reconcile this fact with our findings? The answer to this question came when we considered mutation rate itself as a selective trait and found that when evolutionary runs started initially with higher mutation rate, organisms quickly evolved to high fitness with their proteins both stable and functionally interacting. However, after a certain level of fitness had been achieved, the mutation rate in the population started to gradually yet steadily decrease to a level comparable to modern wild-type bacteria of about $10^{-3}$ mutations per genome per generation (see Fig.6). After a low mutation rate had been fixed in the population, species are restored but the population reaches much higher fitness than in evolutionary runs where mutation rate is set low from the beginning. In other words, evolution in our model found spontaneously a ''simulated annealing'' solution which provided stable species of high fitness. Early in evolutionary runs when proteins have not yet evolved to form proper stable functional complexes, mutations which increase the stability and functional binding are available, making higher mutation rates beneficial. However as the population arrives at local fitness peak(s) so that further point mutations are not capable of producing fitness increase, the mutation rate starts to decrease in order to minimize detrimental mutations in the absence of readily available beneficial ones.  Selection of mutation rates was studied in the framework of phenomenological population dynamics models (*29-33*).  In particular, Gerrish et al recently studied a model where the distribution of fitness effect of a mutation is constant and found that in this case populations tend to increase their mutation rates indefinitely (*32*) . In contrast, our study does not assume a'priori any fitness effect of a mutation – rather the consequences of a mutation are evaluated directly from its impact on protein stability and interactions. We find an almost universal tendency to decrease mutation rate after a certain plateau in fitness is achieved. (A tiny fraction of runs did not result in a dramatic change of the mutation rate.)

Previous studies which employed models of sequence-based genotype-phenotype relationships mostly (but not always, (*34*)) considered organisms represented by a single RNA molecule whose fitness is determined by its secondary structure (*18-20, 35*) in the spirit of the single-fitness peak assumption: all RNA sequences having a particular secondary structure have the same high fitness while any sequence of an alternative secondary structure had low (negligible) fitness. Under such an assumption, the neutral networks naturally appeared and evolutionary dynamics was presented as a drift on such neutral networks (*20, 36, 37*). The genotype-phenotype relationship in this RNA folding model is trivially degenerate (because each pair can switch without affecting other pairs, due to the absence of tertiary interactions). Not surprisingly, delocalization in genotypic space occurs at infinitesimally small mutation rates while phenotypic error catastrophe (when dominant secondary structure is lost) takes place at higher mutation rates (*19, 35*) . Our approach is different in several important respects. First and foremost, we do not make any ''ad hoc'' assumptions that one structure or one sequence dominates the fitness landscape, as we discussed in detail above. Second, our model of protein stability and interactions realistically considers three-dimensional structures of proteins. To this end, any mutation affects stability or interaction – excluding a strictly neutral network scenario. Our findings therefore differ from the RNA secondary structure folding model in several important respects.  In our model, genotypic delocalization occurs only after certain mutation rates are reached, below which the population forms well-defined species. We do not find a ''phenotypic'' delocalization at high mutation rates as native protein structures do not change in simulations, even at high mutation rates (but still below the mutational meltdown threshold (*7*)).

Our model is still minimalistic and it can be improved in a number of ways. It disregards a spatial aspect of the problem where new species can separate in space. Further, it treats variable mutation rate as a continuous variable, while in modern organisms it may only change by discretely switching between wild type and mutator phenotypes.  The analysis of these and other factors within a sequence-based microscopic model of evolution and adaptation is the subject of further work.

**Methods**

Simulations start from a population of 500 identical organisms (cells) each carrying 3 genes with initial sequences designed to be stable in their (randomly chosen) native conformations with $P_{nat} > 0.6$. At each time step, a cell can divide with probability $b$ given by Eq.(1). A division produces two daughter cells, whose genomes are identical to that of mother cells apart from mutations which occur upon replication with probability (mutation rate) $m$ per gene. If any protein in the cell loses its stability ($P_{nat} < 0.6$) by mutation, the cell is discarded. The death rate, $d$, of cells is fixed to 0.005/time unit, and the parameter $b_0$ is adjusted to set the initial birth rate to the fixed death rate ($b=d$). The control coefficient $\alpha$ in Eq.(1) is set to 100. We simulated a chemostat regime: when the population size exceeded 5000 organisms, the excess organisms were randomly culled to bring the total population size to 5000. Initially expression levels are set equally for each protein at $C_i = 0.1$. The expression levels $C_i$ are inherited but they can fluctuate (implicitly modeling phenotypic changes and also mutations of TF proteins and regulatory regions)—at each time step the value of $C_i$ can change with probability 0.01. The change of $C_i$ follows a Gaussian distribution whose mean and standard deviation are 0 and 0.1, respectively, as does the change of mutation rate in Eq.(S5) below. All molecular properties of individual proteins and their interactions are determined directly from genome sequences.

The concentration of free proteins $F_i$ is determined from the Law of Mass Action:

$$F_i = \frac{C_i}{1 + \sum_{j=1}^{3} \frac{F_j}{K_{ij}}} \quad \text{for } i = 1, 2, 3 \quad (2)$$

where $K_{ij}$ is the binding constant of interactions between protein $i$ and protein $j$ (38) and concentrations of binary complexes between all proteins are given by the Law of Mass Action relations:

$$F_{ij} = \frac{F_i F_j}{K_{ij}} \quad (3)$$

We determined, after each change (a mutation or a fluctuation in $C_i$), all necessary quantities by solving the LMA Eqs.(2,3) to find $F_1$ and $F_{23}$ and evaluate the new $P_{nat}$ for mutated protein(s) and $P_{int}^{23}$ for the complex of proteins 2 and 3 to be in their specific binding conformation as explained above. Due to the nonlinear feature of the coupled LMA equations, we use an iterative method here. Once $C_i$ changes, the old set of $F_i$ is substituted into the right side of Eq.(2) and a new set of $F_i$ is calculated. This procedure iterates until the difference between sequential values of $F_i$ drops below the criteria of 1% of the new value.

The following protocol is implemented to model the variable mutation rate. At each time step each organism has a probability $\mu = 0.005$ to change its mutation rate by a random factor:

$$m' = m(1+\varepsilon) \qquad (4)$$

Where $\varepsilon$ is a Gaussian distributed random quantity with mean 0 and standard deviation of 0.1. Upon cell division the mutation rate is inherited by both daughter cells.

Acknowledgements

We thank Konstantin Zeldovich for help at the intial stage of this work and Claus Wilke for useful correspondence. This work is supported by the NIH.

**Model**

In our model, organisms carry 3 genes whose sequences and structures are explicitly represented. Each gene has 81 nucleic acids. Once it is expressed into protein, it folds into a 3x3x3 compact lattice structure. We reduce the range of all possible 3x3x3 lattice structures, which total 103,346, to randomly chosen 10,000 structures for faster calculation. $P_{nat}$ is the probability that the protein folds into the native structure whose energy is the lowest out of 10,000 structures. There exist 144 rigid docking modes between two 3x3x3 lattice proteins, considering 6 surfaces for each protein and 4 rotations for each surface pair of two proteins (6x6x4). $P_{int}^{ij}$ is the probability that two proteins $i$ and $j$ form a stable dimeric complex in the correct docking mode. $P_{nat}$ and $P_{int}^{ij}$ are proportional to the Boltzmann weight factors of the native structure energy, $E_0$, and the lowest binding energy, $E_0^{ij}$ as follows:

$$P_{nat} = \frac{\exp[-E_0/T]}{\sum_{i=1}^{10000}\exp[-E_i/T]}, \quad P_{int}^{ij} = \frac{\exp[-f \cdot E_0^{ij}/T]}{\sum_{k=1}^{144}\exp[-f \cdot E_k^{ij}/T]}. \quad (S1)$$

The binding constants $K_{ij}$ between proteins $i$ and $j$ are calculated as follows:

$$K_{ij} = \frac{1}{\sum_{k=1}^{144}\exp[-f \bullet E_k^{ij}/T]}, \quad (S2)$$

and these values are substituted into the LMA equations (S3) and (S4) to determine the free concentrations of proteins $F_i$ and concentrations of their complexes $F_{ij}$.

We use the Miyazawa-Jernigan pairwise contact potential for both protein structural and interaction energies (1), but scale protein-protein interactions by a constant factor, $f = 1.5$. We chose environmental temperature $T = 0.85$ in Miyazawa-Jernigan potential dimensionless energy units.

**Calculation of Sequence Entropy.**

In order to analyze the degree of diversity of organisms' proteins, we calculated the sequence entropy of these proteins. The sequence entropy of a residue in the $k$-th position is defined as the following (2):

$$S_k = -\sum_{i=1}^{20} P_i^k \log P_i^k, \quad (S3)$$

where $P_i^k$ is the frequency of amino acid of type $i$ in the $k$-th position in a multiple sequence alignment among all organisms in the population. The sequence entropy for a whole protein is obtained by averaging over all 27 positions in its sequence.

**Fitness in Control Simulations.**

In order to determine the role of PPI in shaping the fitness landscape, we carried out a control simulation where fitness – cell growth rate – was determined by stabilities of its proteins only, i.e.

$$b_{control} = b_0 \frac{P_{nat}^1 P_{nat}^2 P_{nat}^3}{1+\alpha(C_1+C_2+C_3-C_0)^2}. \quad (S4)$$

The mutation rate for control simulations were selected in such a way that effective number of mutations per unit time per genome $m \cdot b$ was the same as the low-m case for the full model shown in Fig.2A.

**References**

1. Miyazawa, S. & Jernigan, R. L. (1996) Residue-residue potentials with a favorable contact pair term and an unfavorable high packing density term, for simulation and threading. *J Mol Biol* **256,** 623-44.
2. Yano, T. & Hasegawa, M. (1974) Entropy increase of amino acid sequence in protein. *J Mol Evol* **4,** 179-87.